\documentclass[aps,pre,a4paper,twocolumn,amsmath,amssymb,amsfonts ]{revtex4-1}

\usepackage{graphicx}
\usepackage{epsfig}
\usepackage[colorlinks=True, citecolor=blue, linkcolor=blue, urlcolor=blue]{hyperref}

\newlength{\figwidth}
\renewcommand{\vec}[1]{\mathbf{#1}}
\newcommand{\fig}[1]{Fig.~\ref{#1}}
\newcommand{\eq}[1]{Eq.~(\ref{#1})}

\begin{document}
\title{Scaling of the dynamics of flexible Lennard-Jones chains}
\author{Arno A. Veldhorst}
\email{a.a.veldhorst@gmail.com}
\author{Jeppe C. Dyre}
\author{Thomas B. Schr{\o}der}
\email{tbs@ruc.dk}
\affiliation{DNRF Centre ``Glass and Time'', IMFUFA, Dept. of Sciences, Roskilde University, P.O. Box 260, DK-4000 Roskilde, Denmark}
\date{\today}

\begin{abstract}
The isomorph theory provides an explanation for the so-called power law density scaling which has been observed in many molecular and polymeric glass formers, both experimentally and in simulations. Power law density scaling (relaxation times and transport coefficients being functions of  $\rho^{\gamma_S}/T$, where $\rho$ is density, $T$ is temperature, and $\gamma_S$ is a material specific scaling exponent) is an approximation to a more general scaling predicted by the isomorph theory. Furthermore, the isomorph theory provides an explanation for Rosenfeld scaling (relaxation times and transport coefficients being functions of excess entropy) which has been observed in simulations of both molecular and polymeric systems. Doing molecular dynamics simulations of flexible Lennard-Jones chains (LJC) with rigid bonds, we here provide the first detailed test of the isomorph theory applied to flexible chain molecules. We confirm the existence of isomorphs, which are curves in the phase diagram along which the dynamics is invariant in the appropriate reduced units. This holds not only for the relaxation times but also for the full time dependence of the dynamics, including chain specific dynamics such as the end-to-end vector autocorrelation function and the relaxation of the Rouse modes. As predicted by the isomorph theory, jumps between different state points on the same isomorph happen instantaneously without any slow relaxation. Since the LJC is a simple coarse-grained model for alkanes and polymers, our results provide a possible explanation for why power-law density scaling  is observed experimentally in alkanes and many polymeric systems. The theory provides an independent method of determining the scaling exponent, which is usually treated as a empirical scaling parameter.
\end{abstract}
            
 \maketitle                                      

%\begin{tocentry}
%  \includegraphics[width=\textwidth]{ljc10_cd_im5_dynamics.eps}\\
%  TOC graphic showing the invariance of the dynamics on the isomorph.
%\end{tocentry}

\section{Introduction}
When a liquid or polymer melt is (super)cooled towards the glass transition, its viscosity and relaxation time increase with many orders of magnitude over a relatively small temperature range. More generally, the dynamics of a viscous liquid depends on two variables, density $\rho$ and temperature $T$ (or pressure and temperature). Understanding what exactly controls the viscous slowing down upon cooling and/or compression remains one of the main challenges related to the glass transition~\cite{Debenedetti2001, Dyre2006, Ediger2012}.

An indication that a single, underlying quantity determines the viscous slowing down of supercooled liquids was published in 1998 by T{\"o}lle et al.~\cite{Tolle1998, Tolle2001}. They showed that the dynamics of \textit{ortho}-terphenyl, measured at different densities and temperatures, collapses on a single curve when plotted against a function of density over temperature $h(\rho)/T$. More specifically, these neutron scattering data were found to collapse for $h(\rho)=\rho^4$. Later, a similar scaling was found to work for other organic glass formers, including polymers, showing that the relaxation time is a function of $h(\rho)/T$~\cite{Alba-Simionesco2002, Dreyfus2003, Paluch2003}. There was some debate over the functional form of $h(\rho)$ and whether it could be uniquely determined given the limited density changes experimentally available~\cite{Tarjus2004, Casalini2004, Roland2004, Tarjus2004b, Alba-Simionesco2004, Dreyfus2004}. In a famous review Roland et al.~\cite{Roland2005} demonstrated that scaling with $h(\rho)=\rho^{\gamma_S}$ with a material specific scaling exponent $\gamma_S$ works well for a large group of organic glass formers, including polymers. We refer to this scaling as power-law density scaling. To date, many more molecular liquids have been shown to obey power-law density scaling to a good approximation, including polymers, but also ionic liquids~\cite{Roland2006, Paluch2010, Habasaki2010, Lopez2011, Ribeiro2011, Swiety-Pospiech2012, Swiety-Pospiech2013} and liquid crystals~\cite{Urban2005, Urban2007, Roland2008, Urban2011, Urban2011b, Satoh2013}.

%The dynamics of many organic glass formers, quantified by some relaxation time or the viscosity, can be scaled onto a single curve which is a function of  (power-law density scaling). 
The recently developed isomorph theory\cite{paper4} explains and generalizes power-law density scaling. The isomorph theory predicts that liquids which obey the theory have curves (isomorphs) in their phase diagrams along which structure and dynamics are invariant in the appropriate units. The isomorphs are identified by $h(\rho)/T$ being constant on an isomorph, where $h(\rho)$ is a material specific function. Consequently relaxation times and transport coefficients are  predicted~\cite{Bohling2012} to be functions of $h(\rho)/T$. For sufficiently small density changes  $h(\rho)$  may be approximated by a power law: $h(\rho)\propto\rho^\gamma$, which is equivalent to power law density scaling. In typical experiments, it is possible to change density around 5\%, but recently it has been shown in experiments that $h(\rho)$ is not well approximated by a power law for larger density changes of up to 20\%~\cite{Bohling2012}. Moreover, the theory provides an independent method of determining the scaling exponent $\gamma_S$ for a small density range.
%The predicted invariance of not only relaxation times, but the whole time/frequency dependence of the dynamics explains the experimnetally observed ``isochronal superposition'', i.e. that the shape of e.g., dielectric spectra for some liquids only depends on the relaxation time [REFS]. 
Other predictions of the theory are that certain thermodynamical quantities including the  excess entropy and isochoric specific heat are invariant on the isomorph.  Since both excess entropy and the relaxation times are predicted to be constant on an isomorph, the isomorph theory provides an explanation for Rosenfeld's excess entropy scaling~\cite{Rosenfeld1977, Dzugutov1996, paper4}, according to which a liquid's relaxation times and transport coefficients are functions of excess entropy only.

The isomorph theory has so far only been tested in detail for atomic systems~\cite{paper4, Veldhorst2012}, and for some small rigid molecules~\cite{Ingebrigtsen2012b}. However, many organic glass formers are large molecules or have bulky side groups, because this makes it harder for the liquid to crystallize. These larger molecules, and polymers in particular, inherently have intra molecular degrees of freedom that influence the liquid structure and dynamics. Here, we aim to bridge the gap between the simple models already shown to obey they isomorph theory, and larger flexible glass formers shown experimentally to obey power law density scaling.

 Since both alkanes~\cite{Pensado2008, Galliero2011, Lopez2011} and polymers~\cite{Roland2005, Roland2010} have been shown to obey power-law density scaling, we simulated a general viscous model liquid of linear, flexible Lennard-Jones chains (LJC). The model has been used extensively for viscous polymer melts close to the glass transition~\cite{Bennemann1999d, Binder2003, Riggleman2009, Riggleman2010, Shavit2013}. We show that the LJC liquid has isomorphs in its phase diagram, and we study the effect of the intra molecular degrees of freedom on the applicability of the isomorph theory.

In section~\ref{sec:theory} we give a short overview over the relevant aspects of the isomorph theory. We explain the LJC model in section~\ref{sec:model} and present the details of our simulation method. We start our discussion of the results by showing how the isomorphs were obtained for the LJC model (section~\ref{ssec:making_isomorphs}). We then verify that the dynamics (section~\ref{ssec:dynamics}) and some aspects of the structure (section~\ref{ssec:structure}) are invariant on the isomorph. As predicted by the isomorph theory, we show in section~\ref{ssec:scaling} that isomorph scaling can be used to collapse the dynamics along different isochores onto a single master curve.

\section{Isomorph Theory \label{sec:theory}}
An isomorph is a curve in the phase diagram that consists of state points that are isomorphic to each other. If one takes two state points with $(T_1,\rho_1)$ and $(T_2,\rho_2)$, then pairs of microconfigurations exist with the same coordinates when scaled with density
\begin{equation}
  \rho_1^{1/3}\vec{R}_1 =  \rho_2^{1/3}\vec{R}_2\,.
\end{equation}
Here $\vec{R}=\{\vec{r}_1,\ldots,\vec{r}_N\}$ denotes the coordinates of all particles. 
Two state points are now defined to be isomorphic if these pairs of microconfigurations have proportional Boltzmann weights~\cite{paper4}:
\begin{equation}
  \exp\left(-\dfrac{U(\vec{R}_1)}{k_BT_1}\right) = C_{1,2}\exp\left(-\dfrac{U(\vec{R}_2)}{k_BT_2}\right)\,,
\end{equation}
with $C_{1,2}$ being a proportionality constant that is the same for all physically relevant pairs of microconfigurations, depending only on the two state points. Thus, if two state points are isomorphic, they have the same probability distributions of their reduced unit configurations. From this definition it can be shown that various dynamical and structural properties are invariant on an isomorph, as well as the excess entropy~\cite{paper4}. It should be noted that our model system is expected to only obey the isomorph definition approximately, since the rigid bonds in the molecule do not scale with density. Therefore, the equilibrium configurations at different densities in general are not the same.

The development of the isomorph theory was preceded by the discovery that some liquids have strong correlations in the equilibrium fluctuations of the configurational parts of their energy and pressure. The correlations can be quantified by the standard correlation coefficient~\cite{Pedersen2008, paper1}
\begin{equation}
  \label{eq:R}
  R = \frac{\left< \Delta W \Delta U \right>}
           {\sqrt{\left<(\Delta W)^2\right> \left<(\Delta U)^2\right>}}\,,
\end{equation}
where $U$ is the potential energy, $W$ is the virial, $\Delta$ denotes deviation from thermal average, and brackets $\left<...\right>$ denote average in the canonical ensemble. For liquids where the pair potential is an inverse power law (IPL),  $\upsilon(r) \propto r^{-n}$, the correlation is perfect ($R=1$), but a large group of liquids have a correlation coefficient close to one, indicating strong correlation. Liquids with a correlation coefficient larger than 0.9 were referred to as ``strongly correlating'', but since this term was often confused with strongly correlated quantum systems, we now refer to this class of liquids as ``Roskilde-simple'' liquids.

The standard linear regression slope $\gamma$ of the fluctuations is given by
\begin{equation}\label{eq:gamma1}
  \gamma = \frac {\left<\Delta W \Delta U\right>} {\left<(\Delta U)^2\right>}\,.
\end{equation}
It can be shown using the standard fluctuation formulae that this slope is equal to the logarithmic density derivative of the temperature on a curve of constant excess entropy $S_{ex}\equiv S-S_{ideal}$, where $S_{ideal}$ is the entropy of an ideal gas at the same temperature and density~\cite{paper4}
\begin{equation}\label{eq:gamma2}
  \gamma
    = \frac {\left<\Delta W \Delta U\right>} {\left<(\Delta U)^2\right>}
    = \left(\frac{\partial \ln T}{\partial \ln \rho}\right)_{S_{ex}}\,.
\end{equation}
This slope $\gamma$ is equal to the density scaling exponent $\gamma_s$ mentioned in the introduction, as long the change of density is small enough.

One can use the ``slope'' $\gamma$ calculated from the fluctuations to trace out a curve of constant excess entropy in the phase diagram. First, one calculates $\gamma$ at a certain state point (1) with temperature $T_1$ and density $\rho_1$ using \eq{eq:gamma1}. If one then increments density by a sufficiently small amount to density $\rho_2$, it is possible to calculate the temperature $T_2=T_1(\rho_2/\rho_1)^\gamma$ (\eq{eq:gamma2}) that has the same excess entropy at this new density. This can be done many times in an iterative fashion to obtain a set of state points that have the same excess entropy. Since $\gamma$ may change with density, it is necessary to increment density by a sufficiently small amount (e.g., 2\%), which can be checked by comparing the effect of a different density increment.

Here we use this method to trace out curves in the phase diagram with invariant excess entropy and check if the predicted isomorphic invariance of other properties is fulfilled. 
%Because excess entropy is one of the quantities predicted to be invariant, we call such a set of state points an isomorph. Many things are predicted to be invariant on an isomorph, including thermodynamic quantities like excess entropy, entropy and specific heat, but also the structure and dynamics of the liquid. 
It should be noted that the invariance is only predicted to hold when quantities are considered in the appropriate reduced units, e.g., using $\rho^{-1/3}$ as the unit of length, and $k_BT$ as the unit of energy~\cite{paper4, Fragiadakis2011}. We denote reduced units with a tilde.

The isomorph theory predicts ``isomorph scaling'', i.e., that the dynamics is a function of $h(\rho)/T$, where $h(\rho)$ depends on the system~\cite{Ingebrigtsen2012, Bohling2012}. For atomic systems interacting via a pair potential that is the sum of IPL potentials $\upsilon(r) = \sum_n \upsilon_n r^{-n}$, $h(\rho)$ is given by $h(\rho) = \sum_n C_n \rho^{n/3}$, where the constants $C_n$  are the fractional contributions of each term to the heat capacity~\cite{Bohling2012, Ingebrigtsen2012}. This includes for example the celebrated Lennard-Jones potential~\cite{paper5, Ingebrigtsen2012}. For molecular liquids $h(\rho)$ is not known analytically.

\section{Model and simulation method \label{sec:model}}
We performed Molecular Dynamics simulations of flexible Lennard-Jones chains (LJCs) consisting of 10 rigidly bonded segments. Segments in different molecules and non-bonded segments within a molecule interact via the standard LJ potential, cutting and shifting the potential at $2.5\sigma$. We simulated 200 chains in a cubic bounding box with periodic boundary conditions in the NVT ensemble using a Nos\'{e}-Hoover thermostat. For the time step we used $\Delta t=0.0025$, and the time constant of the thermostat was 0.2. The simulations were performed with our RUMD~\cite{RUMD} software utilizing state of the art GPU computing.

The model has been derived from a model by Kremer and Grest~\cite{Kremer1990}, who did not include the attractive part of the LJ potential. Later, the attractive part has usually been included. Short LJ chains of around ten segments have been used extensively to simulate glassy polymer melts~\cite{Bennemann1998, Aichele2003, Puosi2011, Puosi2012}, even though real polymers easily consist of thousands of monomers. The reason for this is threefold. Firstly, the LJC is a coarse-grained model, meaning that a single Lennard-Jones particle may correspond to several monomers. Secondly, increasing the chain length in general increases the total system size, which in turn increases the simulation time. Most importantly, it is often the equilibrium (viscous) liquid that is of interest. Both increasing the chain length and approaching the glass transition increase the equilibration time, meaning that there is always a trade-off between chain length and viscosity~\cite{Glotzer2002b, Barrat2010}.

Often, the neighboring segments in the chain are bonded by a FENE potential, although harmonic springs~\cite{Riggleman2009, Riggleman2010, Shavit2013} and rigid bonds~\cite{Goel2008, Galliero2009, Galliero2011} have also been used. Here, the bond length $l_b=\sigma=1$ was kept constant using the Time Symmetrical Central Difference algorithm~\cite{Toxvaert2009, Ingebrigtsen2010}. Like other constraint algorithms, these bonds contribute to the virial~\cite{AllenTildesley}: $W_{total} = W_{LJ} + W_{constraint}$, but not to the energy.

With our purpose in mind, the model is of special interest since it has already been shown to obey power-law density scaling, using $\gamma$ as an empirical scaling parameter.~\cite{Galliero2011}.  Moreover, the LJC liquid has been shown to obey Rosenfeld's excess entropy scaling~\cite{Goel2008, Galliero2009, Galliero2011, Voyiatzis2013}.

\section{Results and Discussion}
\subsection{Generating isomorphs \label{ssec:making_isomorphs}}
To generate an isomorph, a NVT simulation was performed at a state point $(\rho_0, T_0)$, and the scaling exponent $\gamma$ was calculated using \eq{eq:gamma1}. We then changed density with 0.02 and used equation \eq{eq:gamma2} to find the temperature at the new state point for which the excess entropy $S_{ex}$ is the same. Applying this procedure iteratively we obtain a curve with constant $S_{ex}$. If the model conforms to the isomorph theory, this curve will be an isomorph, i.e., have invariant dynamics and structure in reduced units. Five prospective isomorphs were generated using this procedure with $\rho_0=1.0$ and $T_0=\{0.5, 0.6, 0.65, 0.7, 0.8\}.$

\begin{figure}
  \centering
  \includegraphics[width=\figwidth]{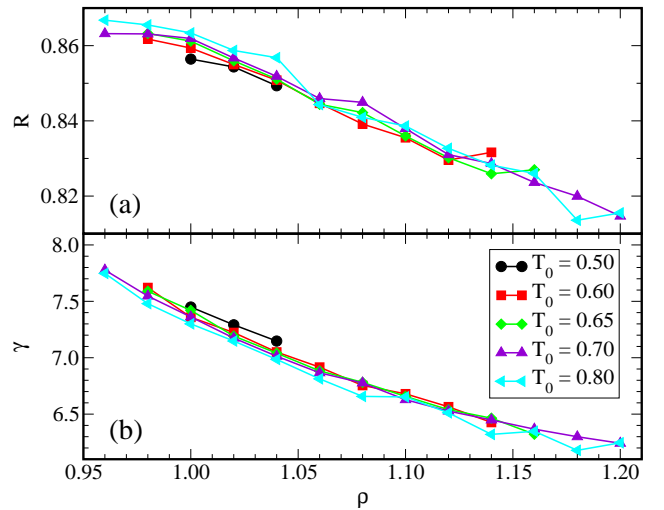}
  \caption{(a) The correlation coefficient $R$, calculated from the instantaneous values of the virial $W$ and the potential energy $U$ using \eq{eq:R}. Each data set corresponds to an isomorph, obtained as described in the text (see \fig{fig:RhoT} for the corresponding temperatures). The correlation coefficient is high, albeit lower than for the single component Lennard-Jones liquid~\cite{paper1, paper2}. (b) The isomorphic scaling exponent $\gamma$ as defined by \eq{eq:gamma1}. The values found are significantly higher than for the single component Lennard-Jones liquids~\cite{paper1, paper2}, and show a clear density dependence. The logarithmic derivatives of $\gamma$ on the isochore and isotherm confirm that $\gamma$ is much more dependent on the density than on temperature: $(\frac{\partial \ln \gamma} {\partial \ln T})_{\rho=1} \approx 0.05$ and $(\frac{\partial \ln \gamma} {\partial \ln \rho})_{T=0.7}\approx0.89 $, as predicted by the isomorph theory.}\label{fig:Rgamma}
\end{figure}

In \fig{fig:Rgamma}(a), the correlation coefficient $R$ is plotted as a function of density for the five isomorphs. For the densities we simulated, the correlation coefficient varies between 0.81 and 0.87, which is lower then the (somewhat arbitrary) 0.9 limit for simple liquids. However, we show with this paper that the LJC model has clear isomorphs in its phase diagram.

In \fig{fig:Rgamma}(b) we plot the values of $\gamma$ calculated from \eq{eq:gamma1}. The isomorph theory predicts $\gamma$ to depend on density but not temperature\cite{paper4,paper5}. This is seen to be fulfilled to a good approximation; $\gamma$ changes much more by increasing density by 25\% than by increasing temperature by 60\%. The density dependence of $\gamma$ means that we can only use \eq{eq:gamma2} for small density changes, which indicates that simple power-law density scaling is an approximation that only works for small density changes.

The $\gamma$ values found for the LJC model (6.1--7.9) are higher than for a single component LJ liquid (5.3--6.7)~\cite{paper1, paper2}. This increase in $\gamma$ is due to the fixed constraints, which can be seen as a very steep repulsion between bonded segments. On the other hand, the high $\gamma$ values is in contrast to the values found from power-law density scaling, which in experiments are generally lower for polymers than for small molecular liquids~\cite{Roland2010}. Tsolou et al~\cite{Tsolou2006} found $\gamma =2.8$ from power-law density scaling of simulation data of a united atom model of \textit{cis}-1,4-polybutadiene. A possible explanation for this low value of $\gamma$ has been given by Xu~\cite{Xu2013} who showed using the generalized entropy theory that polymer rigidity significantly decreases the density scaling exponent $\gamma$. Xu quantified polymer rigidity y the bending energy of the angle between two bonds.

\subsection{Dynamics on an isomorph \label{ssec:dynamics}}
\begin{figure}
  \centering
  \includegraphics[width=\figwidth]{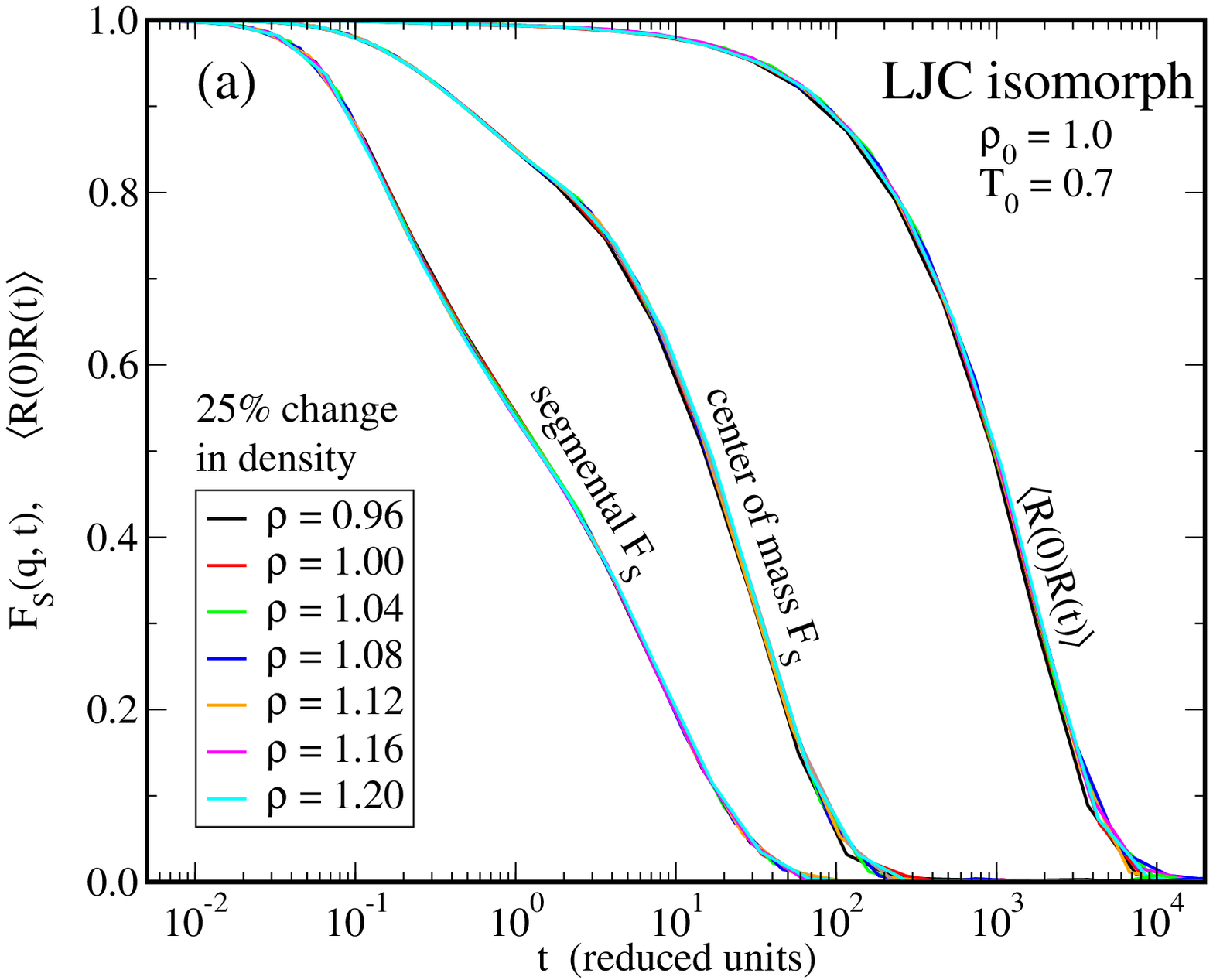}\\
  \includegraphics[width=\figwidth]{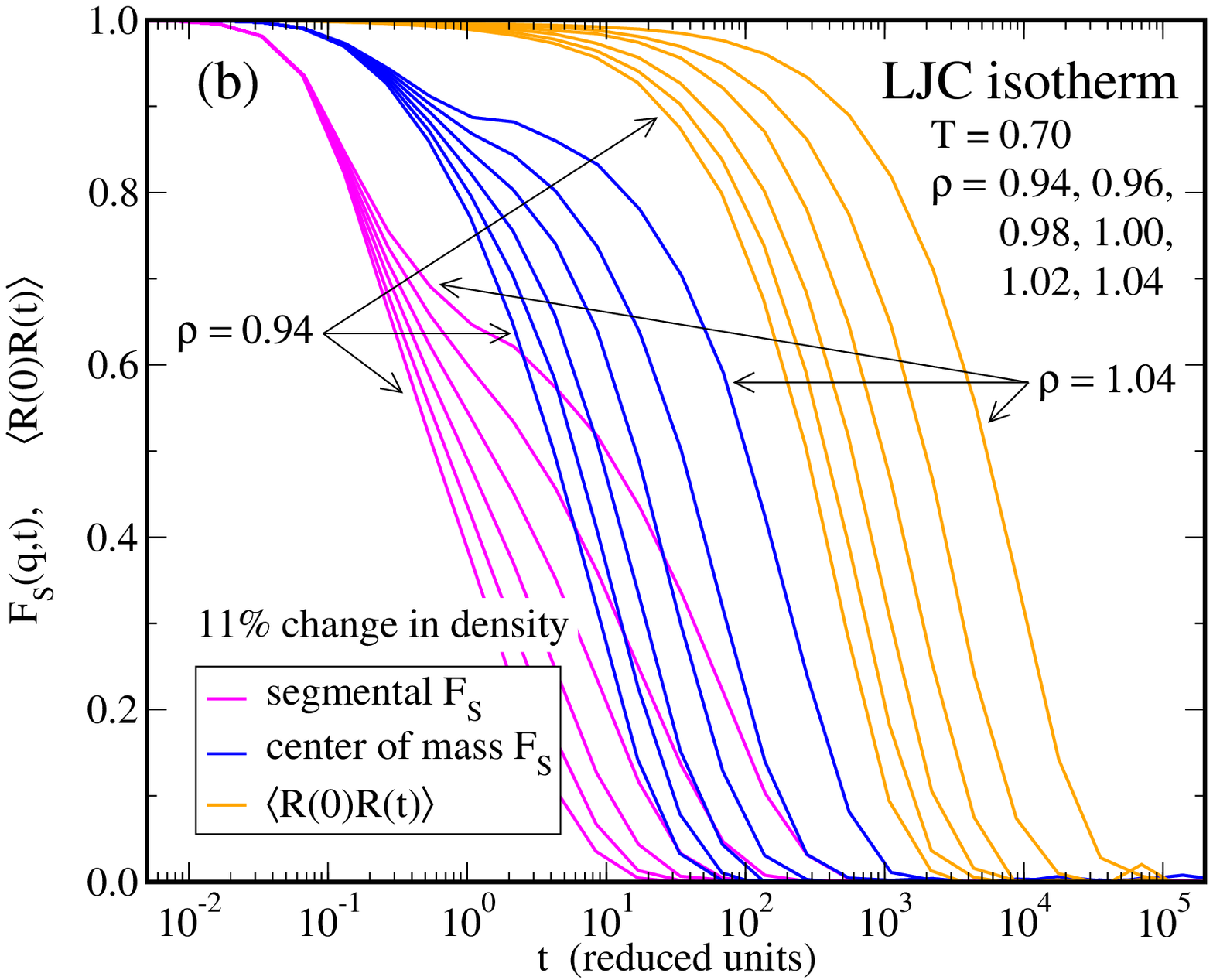}
  \caption{The segmental and center of mass incoherent intermediate scattering function $F_S(q,\tilde{t})$, as well as the normalized orientational autocorrelation function of the end-to-end vector $\left<\vec{R}(t)\vec{R}(0)\right>$. We used $q = 7.09 \rho^{1/3}$ (the position of the first peak of the segmental structure factor). (a) The data for 7 isomorphic state points collapse on a single master curve when plotted in reduced units, and this is the case for all three relaxation functions. (b) For isothermal state points, the curves do not collapse but are spread over a larger dynamical range.}\label{fig:dynamics}
\end{figure}

In the following, we test a number of isomorph predictions focusing on the $(\rho_0, T_0) = (1.0, 0.7)$ isomorph, before returning to the question of the overall scaling properties of the model. The isomorph theory predicts dynamics and structure to be invariant on an isomorph. This invariance applies to data in reduced units, which means that distance and time are scaled using $\tilde{r} = \rho^{1/3}r$ and $\tilde{t} = \rho^{1/3}(k_BT/m)^{1/2}t$, where $m$ is the mass of a segment. The dynamics are of particular interest here, because the dependence on state point becomes large upon cooling and/or compression. In \fig{fig:dynamics}(a), different dynamical quantities are plotted. The self part of the segmental and the center of mass intermediate scattering function $F_S(q,t)$, as well as the normalized orientational autocorrelation of the end-to-end vector $\left<\vec{R}(0)\vec{R}(t)\right>$ are plotted as a function of reduced time. The values of $q$ were kept constant in reduced units: $q=\tilde{q}\rho^{1/3}$ ($\tilde{q} = 7.09$). All these measures of the dynamics collapse well for the isomorphic state points compared to an isothermal density change; Increasing the density by 11\% while keeping temperature constant significantly changes the dynamics, whereas increasing the density 25\% while following the isomorph keeps the dynamics invariant. The data in \fig{fig:dynamics}(a) are in agreement with power law density scaling of segmental and chain relaxation times of simulated polybutadieen~\cite{Tsolou2006}. Our data extend these results by showing that the shape of the entire relaxation curves is invariant.

\begin{figure}
  \centering
  \includegraphics[width=\figwidth]{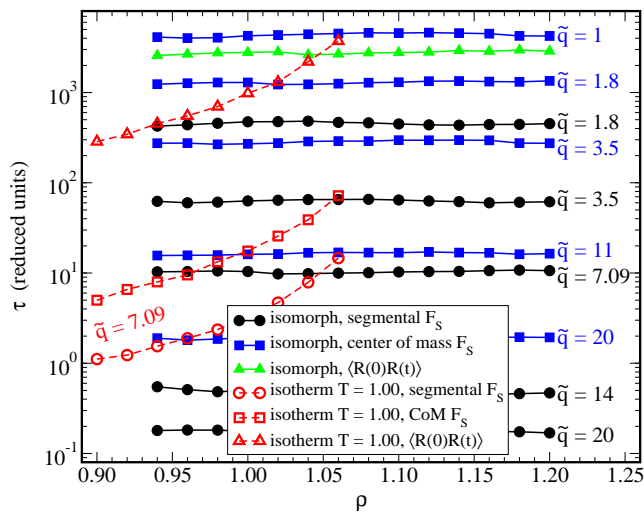}
  \caption{Relaxation times calculated from the orientational autocorrelation of the end-to-end vector and the intermediate scattering function, as function of density. The value of the $\tilde{q}$ vector has been varied to obtain different measures of the relaxation time. Each value was kept constant in reduced units for the different densities All relaxation time measures are invariant for isomorphic state points (filled symbols). An isotherm is included for comparison (open red symbols).}\label{fig:tau}
\end{figure}

We define a relaxation time for the dynamical quantities as the time where the correlation function reaches 0.2. These relaxation times are plotted in \fig{fig:tau}, this time also varying $\tilde{q}$. The different relaxation times characterizing the dynamics covers more than 4 decades in time, but each of them are to a good approximation invariant on the isomorph. In contrast, the relaxation times on the isotherm shown (open red symbols) shows a clear dependence on density. 

\begin{figure}
  \centering
  \includegraphics[width=\figwidth]{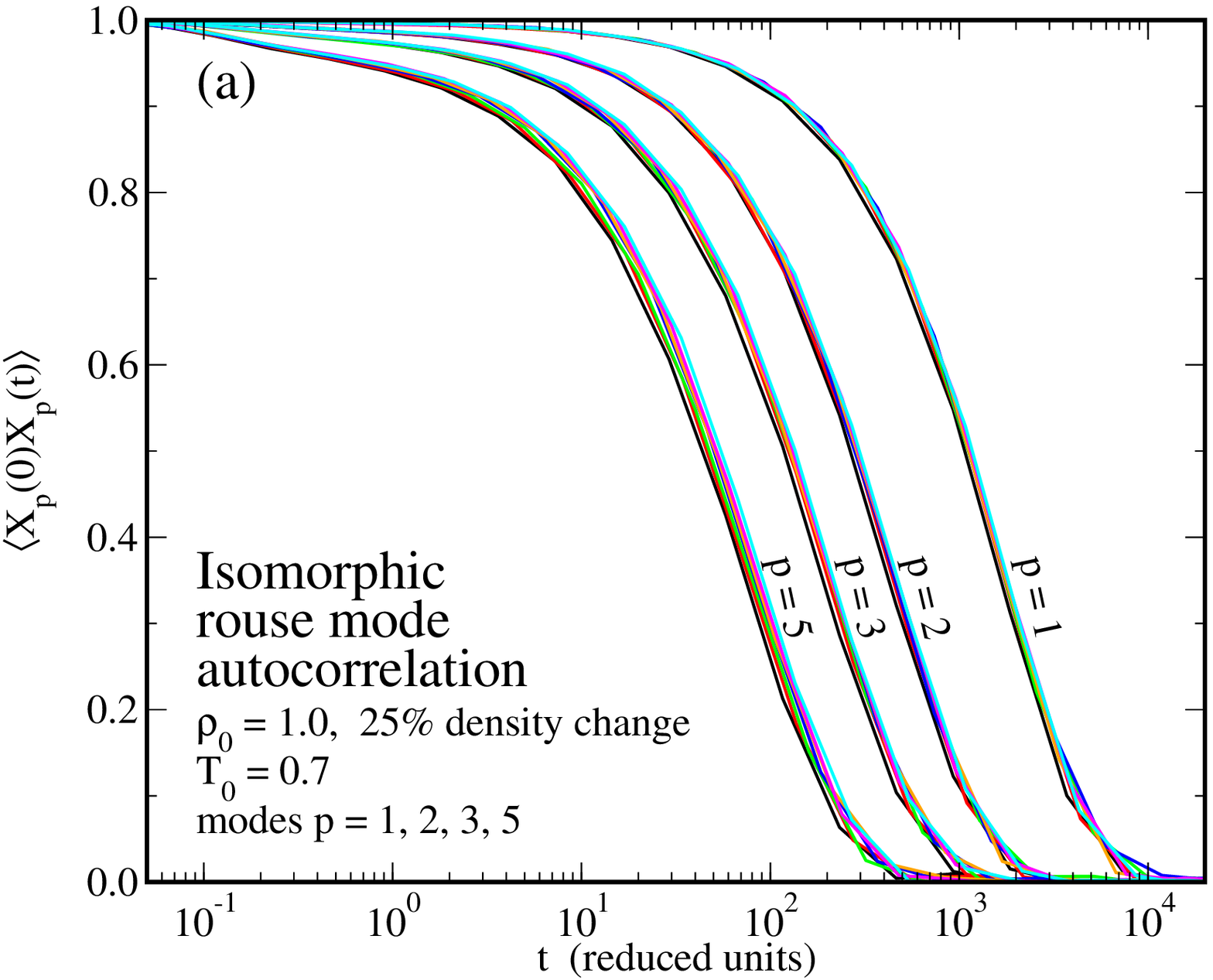}\\
  \includegraphics[width=\figwidth]{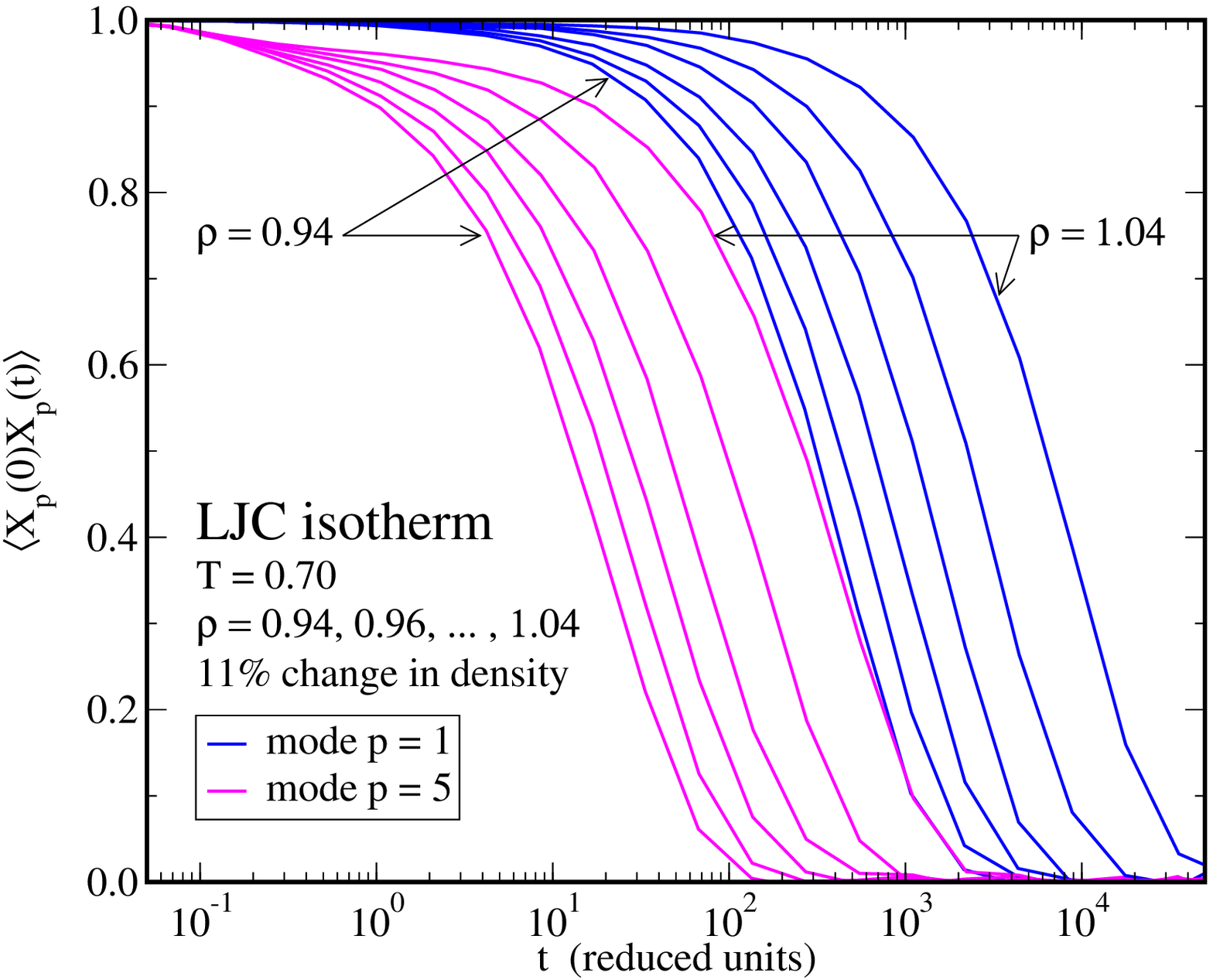}
  \caption{Auto-correlation functions of some rouse modes. (a) For the same isomorphic state points as in \fig{fig:dynamics}(a). The collapse of the Rouse modes is good, especially for the lower modes. (b) Data for the same isothermal state points as in \fig{fig:dynamics}(b). There is no collapse of the dynamics for isothermal state points.}\label{fig:rouse}
\end{figure}

The dynamics of flexible chains are often expressed in terms of correlation functions of Rouse modes, $\left<\vec{X}_p(t)\vec{X}_q(0)\right>$~\cite{Verdier1966, DoiEdwards}. The zeroth mode $\vec{X}_0$ describes the position of the center of mass of the chain, while the higher modes with $p=1,2,\ldots,N-1$ describe the local motion of a subchain of $N/p$ segments. In \fig{fig:rouse} some of the Rouse mode auto correlation functions are plotted for the isomorphic state points. For the lower modes, there is an excellent collapse of the correlation functions, whereas the invariance decreases somewhat for the higher modes. The variance of the highest modes is somewhat surprising considering that the segmental intermediate scattering function shows such a good collapse. It should however be noted that the amplitude of the rouse modes is predicted to scale as $\left<X_p^2\right> \propto 1/(N\sin^2(p/N))$, so the contribution of the higher modes is very small~\cite{Bennemann1999}. Moreover, the $p>0$ Rouse modes represent the conformation of the (sub)chain, and the less than perfect collapse of the highest modes thus indicates that the deviation from isomorph theory is specific to the local intramolecular dynamics. It is well known that reducing the local intramolecular degrees of freedom by including bond and torsional potentials leads to dynamics that are less Rouse-like~\cite{Bernabei2011}. The local degrees of freedom affect mostly the higher modes, giving the standard Rouse behavior for the lower modes representing longer subchains~\cite{Steinhauser2009}

\begin{figure}
  \centering
  \includegraphics[width=\figwidth]{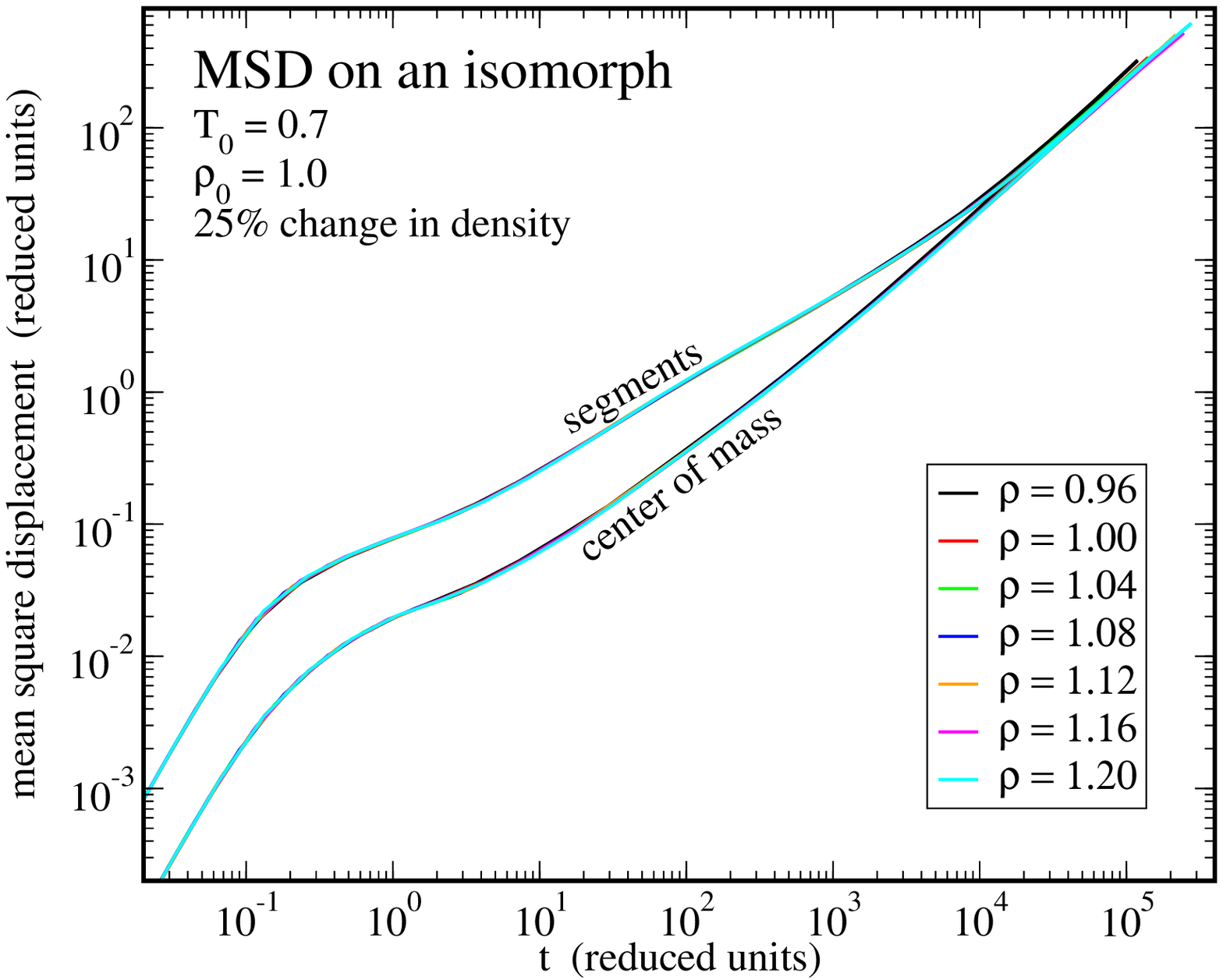}\\
  \includegraphics[width=\figwidth]{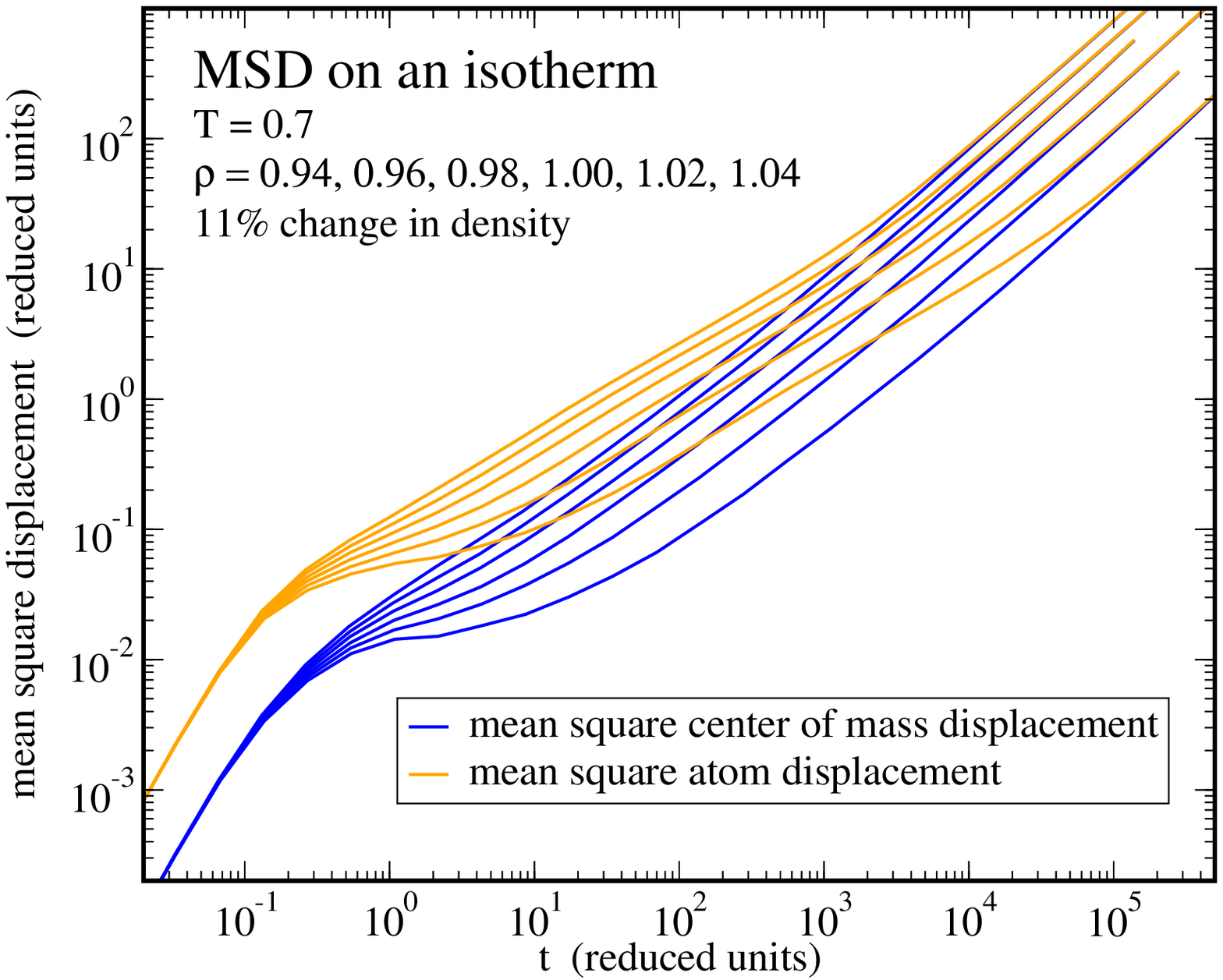}
  \caption{The mean square displacements of the segments and the center of mass of the chains. (a) again, there is a good collapse for the mean square displacement on the isomorph, for both the segments and the center of mass. (b) This is not the case for the isotherm.}\label{fig:msd}
\end{figure}

\fig{fig:msd} shows the isomorphic invariance of the mean square displacement of both the segments and the center of mass in all regime, including the subdiffusive regimes which is specific for polymers and other flexible molecules.

\begin{figure}
  \centering
  \includegraphics[width=\figwidth]{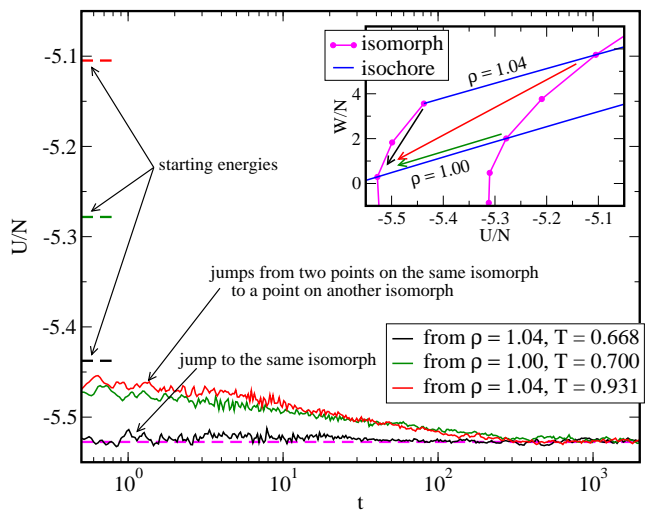}
  \caption{Potential energy relaxation after instantaneous jumps from three different state points to $\rho = 1.00, T = 0.50$. The inset shows the direction of the jumps in the phase diagram, plotted in the $U,W$-plane. Black line: a jump between isomorphic state points. The energy shows no relaxation since the system is immediately in equilibrium. Red and green lines: two jumps from the same isomorph to another isomorph show the same relaxation behavior. The data of the relaxation plots are averages of 8 independent starting configurations.}\label{fig:jump}
\end{figure}

Not only equilibrium dynamics, but also out of equilibrium dynamics is predicted to be invariant on an isomorph. We test this by changing density and temperature instantaneously during a simulation. The center of mass positions of the molecules  are scaled together with the box, but the intramolecular distances were kept constant. In \fig{fig:jump} the relaxation of the potential energy is plotted after different instantaneous jumps. Although the energy is not the same at two isomorphic state points, no relaxation is visible in the energy when jumping between two the to state points (black line). This is predicted by the isomorph theory: two state points on the same isomorph are equivalent with regard to aging~\cite{paper4}. Likewise, when jumping from two state points on the same isomorph to a third state point that is not on that isomorph, the relaxation curve is the same for the two jumps. When the density is changed, the system is immediately in equilibrium at the isomorphic state point with the new density. Any relaxation after the density jump then takes place on the isochore~\cite{Gnan2010}.

\subsection{Structure on an isomorph \label{ssec:structure}}
\begin{figure}
  \centering
  \includegraphics[width=\figwidth]{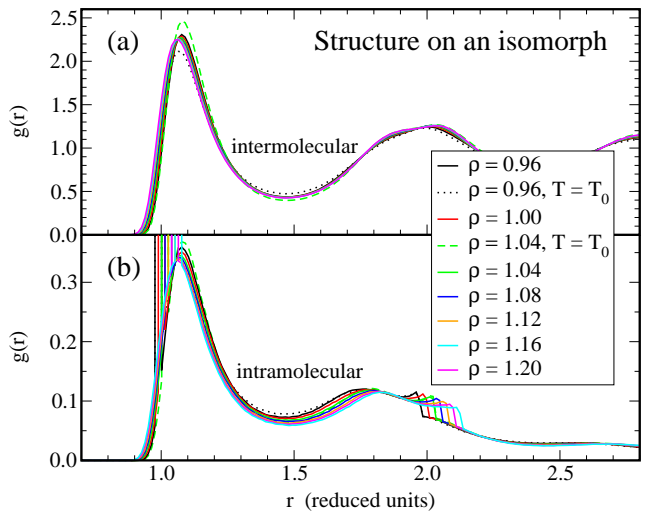}
  \caption{Segmental and molecular structure on an isomorph for the same state points as in \fig{fig:dynamics}. Dashed lines correspond to isothermal density changes and are included for comparison. The data show that intermolecular structure is invariant, while intramolecular structure is not invariant on the isomorph due to the constant bond length. (a) The intermolecular (segmental) radial distribution function $g(\tilde{r})$ in reduced units on the isomorph $(\rho_0, T_0) = (1.00, 0.70)$. The intermolecular $g(r)$ is to a good degree invariant for isomorph state points, especially when compared to a (small) density change on an isotherm. (b) The intramolecular $g(r)$ is clearly not invariant on an isomorph.}\label{fig:rdf}
\end{figure}

Also the structure is predicted to be invariant on an isomorph~\cite{paper4}. However, not all structural quantities are necessarily equally invariant when molecular liquids are considered. Since the length of the rigid bonds is constant in normal units and does not change with density, the bond length in reduced units will not be constant on the isomorph in reduced units. For that reason we plot the inter- and intramolecular contribution to the segmental radial distribution function $g(r)$ separately in \fig{fig:rdf}. The intermolecular structure is quite constant on the isomorph, while the intramolecular structure is clearly not. The center of mass $g(\tilde{r})$ was also found to be invariant on the isomorph when plotted in reduced units (data not shown), but it is also invariant on the isochore and isotherm within the liquid (fluid) phase.

\begin{figure}
  \centering
  \includegraphics[width=\figwidth]{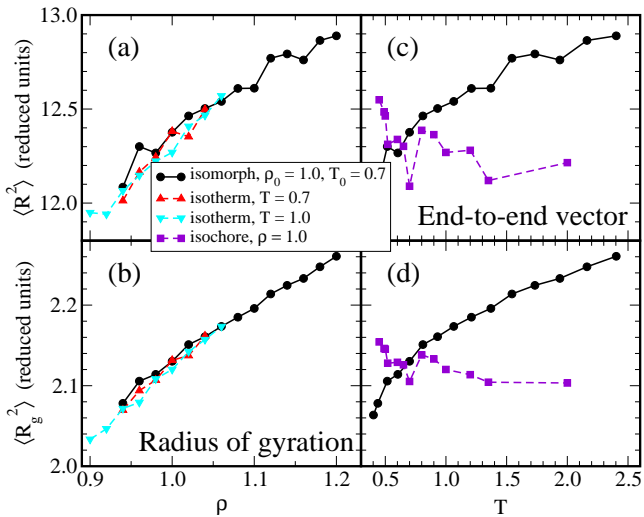}
  \caption{Intramolecular quantities are not invariant on the isomorph (black solid lines) (a) The mean square end-to-end vector $\left<R^2\right>$ and the mean square radius of gyration (b) as a function of density. The temperature dependence of these quantities is similar on the isomorph and isotherms (dashed lines). It should be mentioned that when these quantities are plotted in real units, they show an (intuitive) decrease with density. (c) and (d) The same data for the isomorphic state points, now plotted as a function of temperature and compared with an isochore. These intramolecular quantities are actually more constant on the isochore.}\label{fig:gyradius}
\end{figure}

To investigate the difference in inter- and intramolecular structure further, we plot the mean square radius of gyration $\left<R^2_g\right>$ and the  mean square end-to-end vector $\left<R^2\right>$ in \fig{fig:gyradius}. These intramolecular quantities are clearly not invariant on the isomorph, changing as much with density as on the isotherm. On an isochore these quantities are even more constant than on the isomorph. The lack of temperature dependence of these quantities was already noted for a similar bead-spring model~\cite{Bennemann1999c}.

\subsection{Scaling of the dynamics \label{ssec:scaling}}
\begin{figure}
  \centering
  \includegraphics[width=\figwidth]{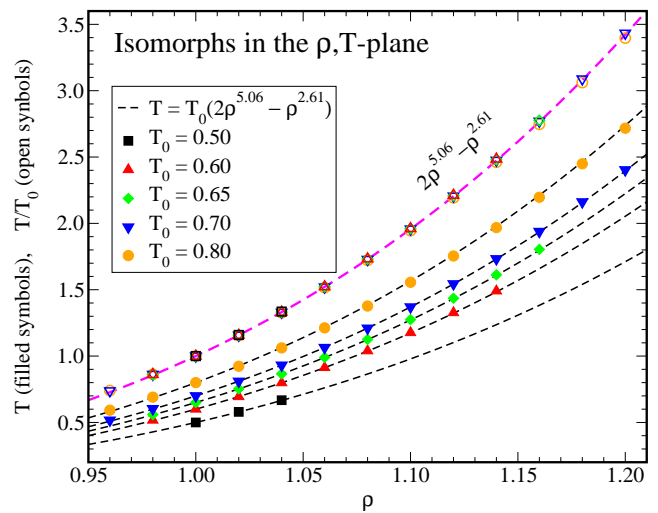}
  \caption{Filled symbols: Shape of isomorphs in the $\rho$,$T$-plane. Open symbols: Same data with temperatures divided by $T_0$, showing a good collapse as predicted by the isomorph theory. Dashed lines: The function $h(\rho)=2\rho^{5.06}-\rho^{2.61}$ was found by fitting to the open symbols (see text).}\label{fig:RhoT}
\end{figure}

Finally, we return to the question of the overall scaling of the dynamics of the model. As mentioned in the introduction, the isomorph theory predicts that each relaxation time characterizing the dynamics is a function of $h(\rho)/T$ where $h(\rho)$ is system dependent function. For atomic systems with pair potentials being sums of power laws, we have an analytical expression for $h(\rho)$~\cite{Ingebrigtsen2012, Bohling2012}. Due to the presence of the bonds, we unfortunately do \emph{not} have an analytical expression for $h(\rho)$ in the model studied here. \fig{fig:RhoT} shows the five studied isomorphs in the $\rho,T$ plane (filled symbols). The open symbols show the same data, except that the temperatures are divided by $T_0$ (the temperature at $\rho=1$). The scaled data is predicted to collapse on a single curve, $h(\rho)$, which is indeed seen to be the case. We have found that the $h(\rho)$ from the single component Lennard-Jones liquid~\cite{Ingebrigtsen2012, Bohling2012} does not describe the shape of Lennard-Jones chain isomorphs correctly due to the rigid bonds (fit not shown). Instead, we have fitted the shape of the isomorphs with a function of the form $h(\rho) = 2\rho^\alpha-\rho^\beta$ where $\alpha$ and $\beta$ are fitting parameters. The choice of the functional form is rather arbitrary; it was found to fit the data well with only two fitting parameters, but there is no a priori reason why $h(\rho)$ should be a sum of two power laws and we do not ascribe any meaning to this functional form. Nonetheless, the shape of the isomorphs is well described by
\begin{equation}
  h(\rho) = 2\rho^{5.06} - \rho^{2.61}\,,
\end{equation}
shown as the dashed pink line in \fig{fig:RhoT}.

\begin{figure}
  \centering
  \includegraphics[width=\figwidth]{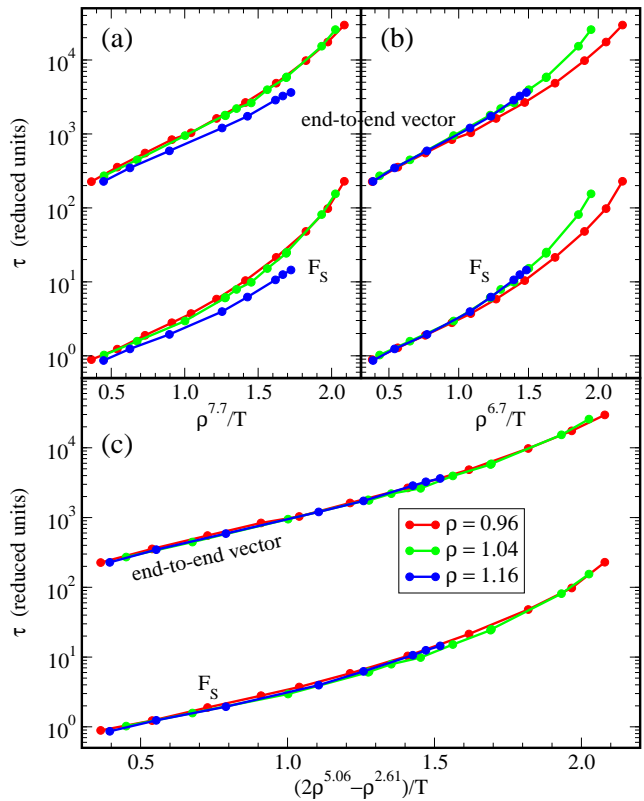}
  \caption{Comparison between power-law density scaling and isomorph density scaling, applied to the relaxation times of the end-to-end vector and the segmental incoherent intermediate scattering function ($F_S$). (a) and (b) The power-law density scaling approach for two different values of $\gamma$ (7.7 and 6.7), collapsing the low and high density isochores respectively. Neither value gives a good collapse of all the data. (c) Isomorph scaling approach, using the function $h(\rho) = 2\rho^{5.06}-\rho^{2.61}$ (see \fig{fig:RhoT}) to scale the relaxation times, giving a much better collapse. %It should be noted that for the lowest density some of the slowest state points have a negative pressure. Although this is of unphysical, it should be noted that the liquid is still homogeneous due to the finite but constant size of the simulation box.
}\label{fig:scaling}
\end{figure}

\fig{fig:scaling} compares for three isochores the power-law density scaling and the scaling predicted by the isomorph theory. \fig{fig:scaling}(a) and \fig{fig:scaling}(b) show that the two smallest densities collapse using power-law density scaling with $\gamma=7.7$, whereas the two highest densities collapse using $\gamma=6.7$. Notice that the values of $\gamma$ found by this empirical scaling is consistent with the values found from the $W,U$ fluctuations in the respective density intervals (see \fig{fig:Rgamma}). The power-law density scaling is an approximation that works well for (relatively) small density changes, and the scaling exponent $\gamma$ can be determined independently from the $W,U$-fluctuations. The more general form of scaling is the one predicted by the isomorph theory, which is tested in \fig{fig:scaling}(c), using the $h(\rho)$ determined empirically in \fig{fig:RhoT}. The collapse is seen to be excellent. Notice that the isomorph scaling also captures the different shapes of the segmental and chain dynamics, which is also well known for power-law density scaling in a small density range~\cite{Roland2004b, Casalini2005, Roland2007}.
% Moreover, it has recently been suggested that for a wider group of glass formers both the $\alpha$-relaxation and the Johari-Goldstein $\beta$-relaxation obey power-law density scaling with the same value of $\gamma$, even though these relaxation times have a completely different dependence on density or $\rho^\gamma/T$~\cite{Ngai2012}. However, the validity this result, especially in the liquid phase, is still under discussion~\cite{Casalini2013}.

\section{Conclusion}
To summarize, we have shown that the predictions of the isomorph theory apply to a flexible chain-like model system, despite the fact that the system is not entirely ``Roskilde-simple'' because the correlation coefficient of the instantaneous $U,W$ fluctuations is less than 0.9. However, the collapse of the dynamics at different time and length scales is unmistakable, and works for the segmental dynamics as well as the chain dynamics. We see a slight deviation from invariance for the highest Rouse modes, but we attribute this to a specific intramolecular effect related to the (local) conformation of the chain. The rigid bonds in the model cannot scale with density and the structure can therefore not be constant on the isomorphs. We have shown that this is only the case for intramolecular structure, while the intermolecular structure stays invariant on the isomorph.

Our results indicate that the isomorph theory may be extended to include flexible molecules. In particular this explains the experimentally observed power-law density scaling for alkanes and many polymers - and predicts that it should break down at larger density variations where isomorph scaling is needed. 

\acknowledgments{The authors are grateful to S{\o}ren Toxvaerd for critical reading of the manuscript. The centre for viscous liquid dynamics ``Glass and Time'' is sponsored by the Danish National Research Foundation via Grant No. DNRF61.
}

%\bibliography{ljchain}

\end{document}